\documentstyle[multicol,aps,prl,epsf,amssymb]{revtex}
\begin{document}
\draft
\title{Flexoelectric surface switching of bistable nematic devices}
\author{Colin Denniston$^1$ and J.M. Yeomans$^2$}
\address{$^1$Department of Physics and Astronomy, The Johns Hopkins
  University, Baltimore, Maryland 21218, USA}
\address{$^2$Department of Physics, Theoretical Physics, University of
Oxford, 1 Keble Road, Oxford OX1 3NP, England.}

\date{\today}
\maketitle

\begin{abstract}
We report on a novel method of dynamically controlling the boundary
conditions at the surface of a nematic liquid crystal using a surface
flexoelectric effect.  By moving the surface directors we show
that one is able to manipulate defects which lie near the surface.
This can be used to produce switching of a nematic liquid
crystal device between two states with very similar free energies.
This results in a bistable device that can retain either state
with {\it no} applied voltage.  Switching between the states occurs when
the movement of the surface directors rotates those in the bulk which
are then able to create or annihilate  defects which lie near the
surface of the device. 
\end{abstract}

\pacs{61.30.Hn,42.79.Kr,61.30.Jf}

\begin{multicols}{2}

Defect structures in liquid crystals naturally evoke interest due to
the appearance of complex textures easily visible by the naked eye.
Early studies focused on classifying the static properties of
the defects and their interactions \cite{CR86,DP93}. More recently the focus
has moved to examining the dynamics of topological defects.

In most practical applications of liquid crystals, such as traditional
display devices, defects destroy the
optical properties and are undesirable.  However novel display
designs, such as bistable displays or multidomain nematics, 
exploit defect properties. Most
attempts to control the defect motion have made use of bulk electric
fields \cite{CK87}.
In this Letter we will show
how the flexoelectric effect can be used to control the surface
alignment of a liquid crystal, thereby manipulating defects near the
surface and switching the state of a liquid crystal display device.

Liquid crystals are typically comprised of highly anisotropic,
rod-shaped molecules.  The nematic phase, which we shall concentrate
on here, occurs when the molecules align parallel giving rise to
long-range orientational order with the direction of alignment
indicated by the so-called director field\cite{DP93}.
In a typical display device, the liquid crystal is confined between
two plates a few microns apart. The director configuration on the
plates is fixed.  When an electric field is switched on the molecules
align in the direction preferred by the field.  After switching off
the field, long-range elastic interactions ensure that the molecules
reorient themselves in the direction preferred by the surfaces. 
These devices can be used as displays because different liquid crystal
orientations have different optical properties.
The switching of such traditional liquid crystal devices is well
understood.

However there is now considerable interest in developing
bistable devices which can retain a memory of two distinct states,
with different liquid crystal orientations, and hence different optical
properties, even when the field is switched off \cite{BB97,BC80}.  It
has been demonstrated, both by minimizing a Landau-de-Gennes free energy 
functional and by experiments\cite{BB97,BT00}, that it is possible to
produce two (meta)stable states with different orientations of the director
in the bulk by having different configurations of defects near the
surface.  However, the process of switching between these two states
is not understood.  As the switching involves changes in the
configuration of the defects near the surface, controlling this
process requires understanding the dynamics of defects and how they
can be manipulated.  

In this Letter we describe the physics behind the switching
dynamics of a simple bistable nematic device. Our main
conclusions are that the driving force which causes the switching is a
surface flexoelectric effect.  This rotates the surface
directors which in turn pull round the bulk director field allowing
the defects which lie near the surface of the device in
one of the bistable configurations to be formed or annihilated. Once
this is done switching is essentially complete and the driving field
can be removed.

The main barrier to understanding the dynamics of liquid crystals
is the complexity of the liquid crystal equations of
motion. The hydrodynamics of liquid crystals are usually described by
the Ericksen-Leslie-Parodi equations\cite{DP93}. However these are not
sufficient to describe switching in bistable devices because they are
restricted to an order parameter of constant magnitude.
Defect motion, which plays a major role in the
bistable switching, is not included. 

Therefore it is necessary to consider a more general formalism of the
hydrodynamics in terms of a tensor order parameter, $\bf Q$\cite{BE94}.
${\bf Q}$  evolves according to a convection-diffusion equation
\begin{equation}
(\partial_t+{\vec u}\cdot{\bf \nabla}){\bf Q}-{\bf S}({\bf W},{\bf
  Q})= {\Gamma} {\bf H}
\label{Qevolution}
\end{equation}
where ${\vec u}$ is the bulk fluid velocity and ${\Gamma}$ 
is a collective rotational diffusion constant.
The term on the right-hand side of Eqn.\ (\ref{Qevolution})
describes the relaxation of the order parameter towards the minimum of
the free energy ${\cal F}$,
 \begin{eqnarray}
{\bf H}&=& -{\delta {\cal F} \over \delta {\bf Q}}+({\bf
    I}/3) {\mbox Tr}\left\{ \delta {\cal F} \over \delta {\bf Q} \right\}.
\label{H(Q)}
\end{eqnarray}  
The order parameter distribution can be both rotated and stretched by
flow gradients.  This is described by the term on the left-hand side
\begin{eqnarray}
{\bf S}({\bf W},{\bf Q})&
=&(\xi {\bf D}+{\bf \Omega})({\bf Q}+{\bf I}/3)+({\bf Q}+
{\bf I}/3)(\xi {\bf D}-{\bf \Omega}) \nonumber \\
&&-2 \xi ({\bf Q}+{\bf I}/3){\mbox{Tr}}({\bf Q}{\bf W})
\end{eqnarray}
where  ${\bf D}=({\bf W}+{\bf W}^T)/2$ and
${\bf \Omega}=({\bf W}-{\bf W}^T)/2$
are the symmetric part and the anti-symmetric part respectively of the
velocity gradient tensor $W_{\alpha\beta}=\partial_\beta u_\alpha$ and
$\xi$ is related to the aspect ratio of the molecules. 
The velocity field ${\vec u}$ obeys
the continuity equation and a Navier-Stokes equation with a stress
tensor generalized to describe the flow of liquid crystals.  The
details of the equations of motion can be found in reference \cite{BE94}.
We stress that the equations include
backflow, allow variations in the magnitude of the order parameter
and follow the hydrodynamics of topological defects. To solve them we
use a recent lattice Boltzmann approach which is proving particularly
robust for problems which relate to complex fluids\cite{DO00}.

We consider a liquid crystal described by the Landau-de-Gennes free
energy \cite{DP93}
\begin{equation}
{\cal F}=\int_V dV \left\{F_b-F_E+F_d \right\}+\int_{\partial V} dS
\left\{F_a\right\},
\end{equation}
where
\begin{eqnarray}
F_b&=&\frac{A}{2} (1-\frac{\gamma}{3})
  Q_{\alpha \beta}^2 - \frac{A \gamma}{3} Q_{\alpha \beta}
Q_{\beta \gamma}Q_{\gamma \alpha}
+\frac{A \gamma}{4}(Q_{\alpha \beta}^2)^2, \nonumber\\
F_E&=&\frac{\epsilon_a}{12 \pi}E_{\alpha}Q_{\alpha\beta}E_{\beta}
\!+\!(\partial_{\beta}Q_{\alpha \beta}) (\epsilon_{f_1}E_{\alpha}+
\epsilon_{f_2}Q_{\alpha \gamma} E_{\gamma}),\nonumber\\
F_d&=&\frac{\kappa}{2} (\partial_\alpha Q_{\beta\lambda})^2,\qquad
F_a=\frac{\alpha_s}{2} (Q_{\alpha \beta}-Q^{0}_{\alpha \beta})^2.
\label{free}
\end{eqnarray}
(Greek subscripts represent Cartesian directions and the usual
summation over repeated indices is assumed.)  The bulk free energy
terms $F_b$ describe a liquid crystal with a
first-order, isotropic--nematic transition at $\gamma=2.7$\cite{DE89}.
Contributions to the free energy which arise from an imposed
electric field $\vec{E}$ are included in $F_E$. Note
that there are bulk terms which depend on {\bf Q} and flexoelectric
terms which couple to the derivatives of {\bf Q}\cite{P91,YN99}.
It is the latter that we shall focus on in this paper. They arise in
liquid crystals where molecules are asymmetric in shape and carry a dipole
moment. Alignment of the dipoles by a field can then lead to a varying
director field being energetically favorable.  $F_d$ describes the
elastic free energy within a one elastic constant approximation\cite{DP93} and
$F_a$ is a surface free energy which fixes a
preferred orientation ${\bf Q}^{0}$ for the surface director field\cite{BC91}.

\begin{figure}
\narrowtext
\centerline{\epsfxsize=3.2in
\epsffile{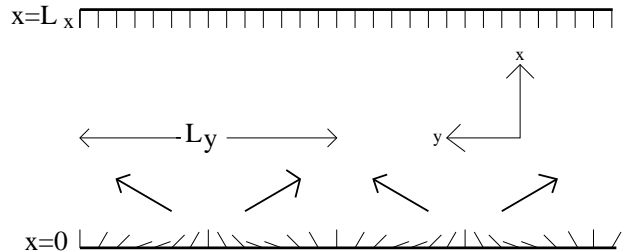}}
\caption{The geometry used to demonstrate flexoelectric switching in a
bistable nematic device. 
In particular note the orientation of the
director on the surfaces of the device.
The bold arrows show the direction in which the surface directors
rotate when a field is applied in the negative $x$-direction.}
\label{schematic}
\end{figure}

We work in two dimensions and consider the geometry shown in
Fig.\ref{schematic}. The device width is $L_x$ and periodic boundary
conditions are imposed in the $y$-direction. The surface pinning
potential at $x=0$ is periodic with the director preferring to lie in the 
$(x,y)$ plane at an angle $\theta_0=(78 \pi / 180) \sin(2 \pi y/L_y)$ to the
$x$-axis. At $x=L_x$ we take $\theta_0=7 \pi /180$. 

Our aim is to show that in a device of this geometry switching can be
driven by a {\em surface} flexoelectric effect and to demonstrate  the kinetic
pathway along which the switching proceeds. Therefore we put
$\epsilon_{a}$ and $\epsilon_{f2}$ in the free energy expression
(\ref{free}) equal to zero\cite{foot1}. Integrating by parts shows
that the remaining flexoelectric term can be rewritten as a surface
contribution 
\begin{equation}
\int_{\partial V}dS\left\{ \sigma_\beta E_\alpha Q_{\alpha\beta} \right\},
\end{equation}
where ${\bf \sigma}$ is the surface normal.  In the geometry of
Fig.~\ref{schematic} we will apply an electric field along $x$.
Symmetry suggests, and we find in our simulations, that the director
remains in the $(x,y)$ plane.  In this case, assuming uniaxiality (so that
$Q_{xx}=q^2(\cos^2\theta-1/3)$ where $q$ is the related to the largest
eigenvalue of $Q$), and ignoring elastic and bulk contributions the
surface free energy density at $x=0$ is
\begin{equation}
F_s= \frac{q^2\alpha_s}{2}(\theta-\theta_0)^2 
+q^2 \epsilon_{f_1} E_x(\cos^2\theta-1/3)
\end{equation}
where $\theta$ is the angle the director makes with the
$x$-axis. Minimising with respect to $\theta$ gives, for small deviations,
\begin{equation}
\theta - \theta_0 = \epsilon_{f_1} E_x/\alpha_s \sin 2 \theta.
\end{equation}

\begin{figure}
\narrowtext
\centerline{\epsfxsize=3.2in
\epsffile{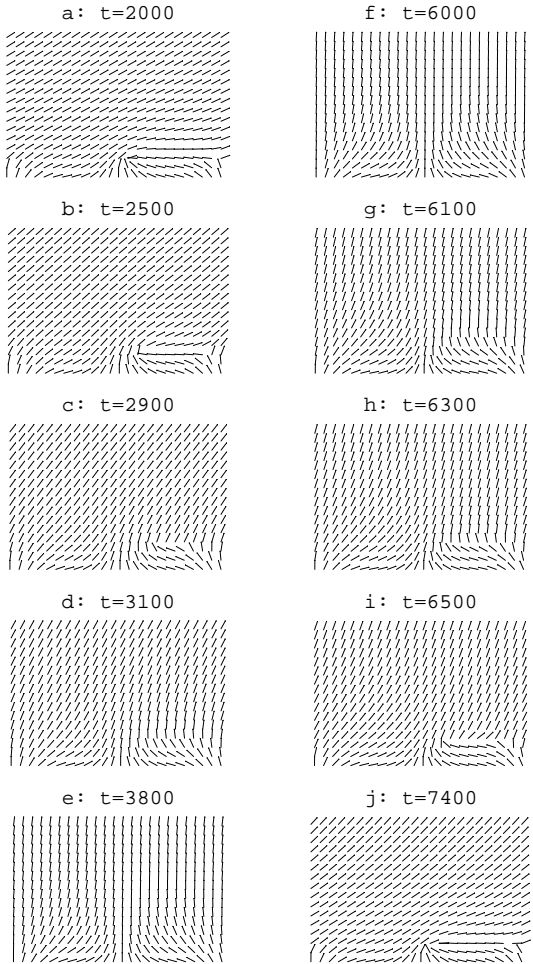}}
\caption{Snapshots of the director configuration during switching: (a)
no field, (b)--(e) switching from diagonal to vertical, (f) no
field, (g)--(j) switching from vertical to  diagonal. Times are
measured in simulation units.}
\label{switch}
\end{figure}

Switching will be driven by a series of voltage pulses
\begin{eqnarray}
V=0,&\;\;\;&0<t<t_0,   \label{pulse1}  \\
V=+V_0,&\;\;\;&t_0<t<t_0+t_1, \label{pulse2}\\
V=0,&\;\;\;&t_0+t_1<t<2t_0+t_1, \label{pulse3}\\
V=-V_0,&\;\;\;&2t_0+t_1<t<2t_0+2t_1. \label{pulse4}
\label{voltage}
\end{eqnarray}
The first voltage pulse corresponds to a field in the negative $x$
direction. Then
for $0<\theta<\pi/2$, $\theta - \theta_0$ is negative; for
$-\pi/2<\theta<0$,  $\theta - \theta_0$ is positive. Therefore
all the surface directors
move towards the vertical as shown by the arrows in Fig.\ref{schematic}. 
This will tend to favour vertical alignment, with the director field
parallel to the $x$-axis, in the bulk.
Conversely a field in the positive $x$-direction causes
the directors to move away from $\theta=0$ favouring a bulk state with the
director field aligned diagonally as shown in Fig.\ref{switch}a.
Thus the surface flexoelectric term can provide
a driving force for the switching.

Simulations of the full Eqns.\ (\ref{Qevolution})-(\ref{free}) displayed in
Fig.\ref{switch} show the path by which switching proceeds.  At each time
we display that portion of the device which lies near the $x=0$ surface.
The voltage pulses cause the system to cycle between the states shown
in Figs. \ref{switch}a and \ref{switch}f which are stable in the zero
field portions of the 
voltage cycle ((\ref{pulse1}) and (\ref{pulse3}) respectively). 
These have very similar
free energies because of the imposed geometry. We find a free energy
difference, measured as a fraction of the elastic energy, of $0.2\%$.
The other frames in Fig.\ 2 show how the system moves between the
bistable states. To switch from diagonal to vertical (Figs.\ 2a--2e)
the surface
deformation pulls the bulk directors towards the vertical. This provides
the driving force for the pairs of defects at the surface to move
together and annihilate. The switch from vertical  to
diagonal (Figs.\ 2f--2j) occurs in a similar way with the bulk rotating
until there is sufficient driving force that the defects can be re-formed.

Further insight into the process can be obtained by considering the
free energy. This is shown in Fig.\ 3 where the bulk, elastic and
total (elastic + bulk + surface + flexoelectric) free energies 
are plotted as a function of time. Upon
switching the field there is a small spike in the free energy which quickly
relaxes as the surface directors move to their new positions. The
switching itself is a far slower process -- the elastic energy
decreases slowly at the expense of the bulk free energy as the
directors are pulled around by the surface
until it is possible to push the defects out of or pull 
them into the device. Defect removal or creation is marked 
by a maximum in the
elastic free energy, which becomes more pronounced as the switching time
becomes longer, together with a faster decrease in the bulk
free energy. Once the elastic free energy has passed this maximum
switching will occur if the field is removed.

The parameters used to obtain the results in Fig.\  3 were, in
simulation units, $A=1.0$, $\Gamma=0.5$, $\gamma=3.0$, $\kappa=2.0$,
$\xi=0.52$, $t_0=2000$, $t_1=2000$,  $L_x=63$,
$L_y=24$. The dimensionless ratio 
$r \equiv \epsilon_{f1} E/ \alpha_s=0.21$ and the voltage was ramped 
up or down over 40 time steps.
The simulation parameters  can be mapped
to physical values for comparison to prototype devices. However
the mapping must be treated with some caution because the simulations
are in two dimensions.  We choose length, time and pressure scales so
that $L_x=1$$\mu$m, pressure$=10^5$Nm$^{-2}$,
viscosities $\sim 0.04-0.1$ Pa.s,
$\kappa=5 \times 10^{-11}$N and  switching
times $\sim 10^{-4}$s.

$r$ is a
measure of the efficacy of the flexoelectric effect in
rotating the surface directors. For a given device this ratio is
controlled by the applied voltage. As expected the switching time
increases rapidly with decreasing $r$.
As $r$ decreases further switching fails. For example, for $r=0.05$ and
$L_x=63$, switching had not occurred after $10 $ms. Moreover there
was no discernible variation of the free energy with time suggesting
that switching would not occur at a later time.

It is also interesting to consider the effect of increasing $L_x$ on switching
times. The vertical to diagonal switching time is little
changed by doubling the device width, whereas the diagonal to
vertical switching time increases substantially. This is because an
increase in $L_x$, without changing other parameters, favours the
diagonal state over the vertical one: the difference in their free
energies is now $5.8\%$ of the elastic free energy.

 As $\kappa$ is decreased the switching time
increases. This happens because  the diffusion constant for the
relaxation of the orientation of the director is proportional to
$\kappa$. For small $\kappa$ switching fails. This occurs when the
elastic coupling between the surface and bulk directors becomes too small
for the bulk to be rotated by a surface perturbation.

\begin{figure}
\narrowtext
\centerline{\epsfxsize=3.0in
\epsffile{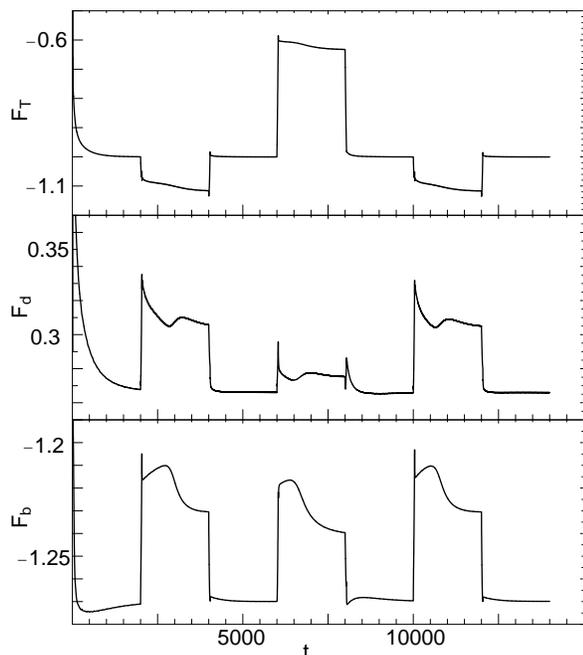}}
\caption{Variation of the total, elastic and bulk free
energies (measured relative to the total free energy with no applied
field) with time t. A field along
$-\vec{x}$ is switched on at $t_0=2000$ and off at $t_0+t_1=4000$.
A field along
$+\vec{x}$ is switched on at $2t_0+t_1=6000$ and off at $2t_0+2t_1=8000$.
Times, measured in simulation units, are comparable in Figs.\ 2 and 3.}
\end{figure}

In the simulations it is possible to ``turn off'' back-flow effects,
that is, to ensure that any movement of the director field does not
induce a flow. This has little effect upon the switching time.
Increasing the temperature towards  that of the nematic--isotropic
transition leads to a decrease in switching time because the defects
have a lower pinning energy. 

The simple device described here has allowed us to obtain a fuller
understanding of a possible switching mechanism in bistable
displays.  Experiments on prototype devices have reported switching times
somewhat longer than those calculated here.  One likely reason for
this is that we have tuned the free energy of the two metastable
states to be very nearly equal.  As we saw when changing the device
width, the switching time is strongly dependent on this free energy
difference.
Moreover, in a real liquid crystal there will in 
general be a bulk field term in
the free energy ($\epsilon_a \neq 0)$. 
This may cause the switching in one direction to be
due to a Frederiks' transition and inhibit switching in the other
direction. A change in boundary conditions could be used to compensate
for this imbalance. Other differences in prototype devices are that the
modulated director structure at 
the surface is created using normal alignment on a blazed surface
rather than modulated alignment on a flat surface. Work to
incorporate this additional physics is currently underway.

To conclude, we have shown that it is possible to model the 
switching mechanism in a bistable nematic device using lattice
Boltzmann simulations of the liquid crystal equations of motion. Thus
we have been able to show that 
switching can be driven by a surface
flexoelectric term in the free energy.  Bistability occurs because the
diagonal and vertical director alignments have very similar free
energies. They remain stable because there is a free energy barrier
between them.
To promote switching, defects need to be
introduced to or expelled from the device: this is most easily
accomplished if the defect pinning energy is small.
~\\
Acknowledgements: It is a pleasure to thank S.J. Elston, N. Mottram, 
E. Orlandini, M.J. Towler, G. Toth and H. Walton for many
helpful discussions. We acknowledge funding from the EPSRC and Sharp
Laboratories of Europe Ltd.  C.D. acknowledges funding from NSF Grant
No. 0083286.


\vfill
\end{multicols}


\begin{references}
\bibitem{foot1}
Preliminary simulations with these parameters nonzero indicate that
the qualitative features of the switching remain the same.
\end{references}
\end{document}